\documentclass[preprint,3p]{elsarticle}
\usepackage{amsmath}
\usepackage{comment}
\usepackage{color}
\usepackage{braket}
\usepackage{amssymb}
\usepackage{makeidx}
\def\Tr{\mathrm{Tr}}

\def\simge{\mathrel{%
    \rlap{\raise 0.511ex \hbox{$>$}}{\lower 0.511ex \hbox{$\sim$}}}}
\def\simle{\mathrel{
    \rlap{\raise 0.511ex \hbox{$<$}}{\lower 0.511ex \hbox{$\sim$}}}}

\newcommand \be{\begin{eqnarray}}
\newcommand \ee{\end{eqnarray}}

\newcommand{\<}{\left<}
\renewcommand{\>}{\right>}
\newcommand{\zb}{\bar{z}}
\newcommand{\wb}{\bar{w}}
\def\XXint#1#2#3{{\setbox0=\hbox{$#1{#2#3}{\int}$}
\vcenter{\hbox{$#2#3$}}\kern-.5\wd0}}

\begin{document}

\begin{frontmatter}



%% e.g. \dochead{17th International Conference on Dynamical Processes in Excited States of Solids}

\title{Squared eigenvalue condition numbers and eigenvector correlations from the single ring theorem}

\author[a1]{Serban Belinschi}
\ead{serban.belinschi@math-univ.toulouse.fr}
\author[a2]{Maciej A. Nowak}
\ead{nowak@th.if.uj.edu.pl}
\author[a3]{Roland Speicher}
\ead{speicher@math-uni.sb.de}
\author[a2]{Wojciech Tarnowski}
\ead{wojciech.tarnowski@uj.edu.pl}

\address[a1]{CNRS, Institut de Math\'{e}matiques de Toulouse, 118 Route de Narbonne, F-31062 Toulouse Cedex 09, France}
\address[a2]{M. Smoluchowski Institute of Physics and Mark Kac Complex Systems Research Centre, Jagiellonian University,  PL--30348 Krak\'ow, Poland}
\address[a3]{Universit\"at des Saarlandes, Fachrichtung Mathematik, Postfach 151150, 66041 Saarbr\"ucken, Germany}
%\cortext[cor1]{Corresponding author.}
\author{}

\address{}

\begin{abstract}

We extend the so-called "single ring theorem"~\cite{SINGLERING}, also known as the Haagerup-Larsen theorem~\cite{HAAGLAR}, by showing that  in the limit when the size of the matrix goes to infinity
a particular correlator  between left and right eigenvectors of 
the relevant non-hermitian matrix $X$, being the spectral density weighted by the squared eigenvalue condition number, is  given by a simple formula involving only  the radial spectral cumulative distribution function of $X$. We show that this object allows to calculate the conditional expectation of the squared eigenvalue condition number. We give examples and we provide cross-check of the analytic prediction by the large scale numerics.

\end{abstract}


\end{frontmatter}

%%%%%%%%%%%%%%%%%%%%%%%%%%%%%%%%%%%%%%%%%%%%%%%%%%%%%%%%%%%%%%%%%%%%%%
%%%%%%%%%%%%%%%%%%%%%%%%%%%%%%%%%%%%%%%%%%%%%%%%%%%%%%%%%%%%%%%%%%%%%%
\section{Introduction}

Recently, Belinschi, Speicher and  {\'S}niady ~\cite{BSS} provided a rigorous mathematical justification of the method of the so-called generalized Green's functions~\cite{SINGLERING,JNPZ,JNPWZ,JNNPZ,CW}, broadly exploited in the physics literature in relation to the spectral problems of non-hermitian 
operators.
 In particular, as one of the examples of their construction, they have shown explicitly, how the generalized  Green's functions reproduce the Haagerup-Larsen theorem~\cite{HAAGLAR}, valid for the case when the non-hermitian operator $X$ can be decomposed as $X=PU$, where $P$ is a positive hermitian operator, $U$ is a Haar unitary operator and $P$ and $U$ are mutually free in the sense of free random variables~\cite{VOICULESCU}. Such operators are named ''$R$-diagonal'' in the mathematical  literature \cite{NicaSpeicher}. It was shown \cite{Guionnet} that random matrices drawn from a probability distribution function of the form $P(X,X^{\dagger})\sim \exp\left(-N\Tr V(XX^{\dagger})\right)$ (biunitarily invariant ensembles) in the limit $N\to\infty$ become $R$-diagonal. 
The Haagerup-Larsen theorem for such $R$-diagonal operators states that the spectrum possesses radial symmetry, is localized within 
%two-rings
the two circles with known radii $r_{\rm min}, r_{ \rm max}$ (including the possibility $r_{ \rm min}=0$, $r_{ \rm max}=\infty$),  and the radial spectral cumulative distribution function $F(r)=2\pi \int_{0}^{r} s \rho(s)ds$ can be derived from the simple functional equation $S_{P^2}(F(r)-1)=\frac{1}{r^2}$, where $S_{P^2}(z)$ is the Voiculescu S-transform for the square of the positive operator $P$ and $r$ is the modulus of the complex eigenvalue $\lambda$.  The Haagerup-Larsen theorem gives the mapping between spectral densities of eigenvalues and singular values for biunitarily invariant ensembles in the large $N$ limit. Recently, the correspondence between eigenvalues and singular values was extended to the exact mapping between their joint probability density functions~\cite{KIEBURGHOSTERS}. In this letter we demonstrate that the function $F(r)$  yields, for these biunitarily invariant ensembles,
also  (in the limit $N \rightarrow \infty$) the eigenvector correlation function, %i.e. 
namely we show that
\be
O(r) \equiv \lim_{N\to\infty} \frac{1}{N^2}\left< \sum_{\alpha} O_{\alpha \alpha}\delta^{(2)}(\lambda-\lambda_{\alpha}) \right> =\frac{1}{\pi}\frac{F(r)(1-F(r))}{r^2},
\label{main}
\ee
where $\<\dots\>$ denotes expectation value,  $O_{\alpha \beta}=\<L_{\alpha}|L_{\beta}\>\<R_{\beta}|R_{\alpha}\>$, where $\left|L_{\alpha}\>$ and  $\left|R_{\alpha}\>$ are left and right eigenvectors of $X$, respectively.  We make use of free probability tools, thus the result is valid at the $N\to\infty$ limit. In case of finite but large size of matrices the formula quite well describes the correlator in the bulk of the spectrum, however the transient phenomena near the spectral edge (typically of size $1/\sqrt{N}$) are not accessible within this formalism.

This paper is organized as follows.  Section~\ref{sec:GGF} recalls the definition of the generalized Green's functions~\cite{JNPZ}. Relevance of the correlation function and its connection with the eigenvalue condition number is discussed in Section~\ref{sec:DiagOverlap}.
Section~\ref{sec:CorrFromHL} exploits the formalism and results of~\cite{BSS}, in order to provide a short, direct proof of the main result (\ref{main}). 
Section~\ref{examples} includes few examples where our formula can be easily applied and provides the verification of these  results with the large scale numerical simulations. We derive in Section \ref{sec:eigenvec-eigenval} a mapping between the spectral density and the eigenvector correlator, showing that they play an equal role in the biunitarily invariant ensembles.
Section~\ref{conclusions} concludes the paper.

\section{Generalized Green's functions} \label{sec:GGF}
\label{quatmeth}

In this section we briefly summarize  the method of the generalized Green's function  for non-hermitian random matrix models  in the limit $N\rightarrow \infty$.
The method is based on the ``electrostatic'' analogy~\cite{STERN,FIODOROV,BROWN}. One defines a quantity
\begin{align}
\label{pot1}
	\Phi(z,\zb,w,\wb) = 
	\frac{1}{N} \left < {\rm Tr} \log \left ( (z {\bf 1}_N-X)(\bar{z}{\bf 1}_N - X^\dagger) + |w|^2 {\bf 1}_N \right ) \right >,
\end{align}
which can be interpreted in the limit $|w| \rightarrow 0$ as an electrostatic 
potential of a cloud of $N$ identical electric charges interacting on the $z$-complex plane. 
The corresponding electric field is a gradient of the potential
\begin{align}
\label{GPhi}
	G\left(z,\zb, w,\wb \right) = \partial_{z} \Phi(z,\zb,w,\wb) = \frac{1}{N}
	 \left < {\rm Tr} \frac{\bar{z}{\bf 1}_N-X^{\dagger}}{(z {\bf 1}_N-X)(\bar{z} {\bf 1}_N-X^{\dagger} )+|w|^{2} {\bf 1}_N } \right >.
\end{align}
We study first  the distribution of eigenvalues
\be
\rho (z,\zb) \equiv \frac{1}{N}\left < \sum_i {\delta^{(2)} 
\left(z - \lambda_{i}\right)}\right >,
\ee
where $\lambda_i$'s are the eigenvalues of $X$. The limiting eigenvalue density comes from   the Gauss law 
\begin{align}
\label{gausslaw}
\rho (z,\zb) =  \lim_{|w| \rightarrow 0} \frac{1}{\pi} \partial_{\bar{z}} G\left(z,\zb,w,\wb\right).
\end{align}
This relation follows from a standard representation of the complex Dirac delta function 
\be
\pi\delta^{(2)}\left(z-\lambda_{i}\right)=\underset{|w| \rightarrow 0}{\lim}\frac{|w|^{2}}{\left(|w|^{2}+\left|z-\lambda_{i}\right|^{2} \right)^{2}}. \label{Diracdelta}
\ee
The expression in the brackets on the r.h.s. of (\ref{GPhi}) can take formally 
 the standard form of the resolvent $(z-X)^{-1}$ 
at  the price of introducing $2N\times 2N$ matrices
\begin{align}
Q \otimes {\bf 1}_N = \left ( \begin{matrix} z {\bf 1}_N & -\bar{w}{\bf 1}_N \\
	 w {\bf 1}_N& \bar{z}  {\bf 1}_N
	 \end{matrix} \right ) \ , \quad  
	 \mathcal{X} = \left ( \begin{matrix} X & 0 \\
	 0 & X^\dagger  
	 \end{matrix} \right ) ,
\end{align}
in place of the original $N\times N$ ones. The generalized resolvent is represented by a  $2\times 2$
matrix 
\begin{align}
\label{quatgf}
\mathcal{G}(z,\zb,w,\wb) \equiv
\left( \begin{matrix} \mathcal{G}_{11} & \mathcal{G}_{12} \\
\mathcal{G}_{21} & \mathcal{G}_{22}  
\end{matrix} \right ) =
\frac{1}{N} \left < {\rm bTr} \frac{1}{Q  \otimes {\bf 1}_N-\mathcal{X}} \right >=\left ( 
 \begin{matrix} \partial_z \Phi & \partial_w \Phi \\
 -\partial_{\bar{w}} \Phi & \partial_{\bar{z}} \Phi
 \end{matrix} \right ),
\end{align}
where the block-trace is defined as
$$
	\rm bTr \left ( \begin{matrix} A & B \\
	 C & D  
	 \end{matrix}\right ) = \left ( \begin{matrix} \Tr A & \Tr B \\
	 \Tr C & \Tr D  
	 \end{matrix}\right ).
$$
We note that  $\mathcal{G}$ from (\ref{quatgf})  has the algebraic structure of quaternions and we refer to it as the
generalized Green's function or the quaternionic  Green's function
\cite{JNPZ,JNPWZ,JAROSZNOWAK}, since both  $\mathcal{G}(z,\zb,w,\wb)$ and $Q$  are quaternions. Similarly, one can define the quaternionic R-transform,  ${\cal R}[{\cal G}(Q)]+[{\cal G}(Q)]^{-1}=Q$, which is additive under the free convolution of non-hermitian ensembles and generates also the non-hermitian multiplication laws~\cite{BJN}.  We mention that the quaternionic extension is equivalent to another approach known under the name of hermitization method \cite{SINGLERING,CW,GIRKO}, in which the diagonal and off-diagonal blocks of matrices $Q$ and $\cal{X}$ are flipped before the block-trace operation.

 The upper-left element of the quaternionic resolvent $\mathcal{G}_{11}$ is equal to $G(z,\zb,w,\wb)$ (\ref{GPhi}), the second diagonal element ${\cal G}_{22}$ is just its complex conjugated copy,  but one may wonder what role is played by  the off-diagonal  elements of the 2 by 2 matrix ${\cal G}$? 
 
 If a non-normal matrix $X$ is diagonalizable via a similarity transformation, it possesses distinct left and right eigenvectors
$X=\sum_{\alpha} \lambda_{\alpha} \left|R_{\alpha}\>\<L_{\alpha}\right|$,
where   $X\left|R_{\alpha}\>=\lambda_{\alpha} \left|R_{\alpha}\>$  and  $\<L_{\alpha}\right|X=\lambda_{\alpha}\<L_{\alpha}\right|$ with the normalization
 $\<L_{\alpha}|R_{\beta}\>=\delta_{\alpha \beta }$. However, neither the left and right eigenvectors are normalized to unity $\<L_{\alpha}|L_{\alpha}\> \neq 1 \neq \<R_{\alpha}|R_{\alpha}\>$ nor they are orthogonal to each other $\<L_{\alpha}|L_{\beta}\> \neq 0 \neq \<R_{\alpha}|R_{\beta}\>$ (for $\alpha \neq\beta$). The orthogonality relations hold only for normal matrices. We define a bra of the right eigenvector in the standard way $\< R_i\right|=\left(\left| R_i\>\right)^{\dagger}$, and analogously a ket of the left eigenvector.
 
   The biorthogonality condition leaves the freedom of rescaling each eigenvector by an arbitrary  non-zero complex number $\left|R_i\right>\to c_i \left|R_i\right>$, $\left<L_i\right|\to\left<L_i\right|c_i^{-1}$. Another allowed transformation is the multiplication by a unitary matrix $\left|R_{i}\right>\to U\left|R_i\right>$, $\left<L_i\right|\to\left<L_i\right|U^{\dagger}$. The simplest non-trivial quantities invariant under these transformations form the matrix of overlaps $O_{\alpha\beta}\equiv \<L_{\alpha}|L_{\beta}\>\<R_{\beta}|R_{\alpha}\>$. One can define a correlation function~\cite{CHALKERMEHLIG} (being the special case of the Bell-Steinberger matrix \cite{BELL,SAVINSOK,FYODSAV})
 \begin{align}
\label{evs}
O(z,\zb) \equiv \frac{1}{N^2} \left < \sum_\alpha O_{\alpha\alpha} 
\delta^{(2)}(z -\lambda_\alpha) \right >.
\end{align}
 
 Below we show that the product of off-diagonal elements of ${\cal G}$ in the limit $|w|\to 0$ gives the eigenvector correlator, simplifying the proof, originally given in~\cite{JNNPZ}.

We rewrite the electrostatic potential \eqref{pot1} in terms of the regularized Fuglede-Kadison determinant ~\cite{KADISON}, using the relation Tr $ \log A= \log \det A $ and linearize it by bringing it to a block structure   
 \begin{align}
\label{potKF}
	\Phi(z,\zb,w,\wb) = 
	\frac{1}{N} \left < {\rm log}  \det \left ( (z-X)(\bar{z} - X^\dagger) + |w|^2 \right ) \right >=\frac{1}{N}\left< {\rm log} \det (Q-{\cal X})\right>.
\end{align}
 
We assume that the matrix $X$ can be diagonalized by a similarity transformation $X=S\Lambda S^{-1}$, which enables us to rewrite the Fuglede-Kadison determinant as 
 \be
  \det (Q -{\cal{X}})=\det [ {\cal S}^{-1}(Q  -{\cal X}) {\cal S}]= \det \left( \begin{array}{cc}
z{\bf 1}_N-\Lambda &-\bar{w}\<L| L\>\\
w \<R|R\> & \bar{z}{\bf 1}_N-\bar{\Lambda}
\end{array}
\right)=\det(A+F),
\ee
where ${\cal S}={\rm diag}(S, (S^{-1})^{\dagger})$ is a block-diagonal matrix. Here $S(S^{-1})$ are built from $N$ right (left) eigenvectors of $X$. In the last equality we represent the matrix as a sum of diagonal $A$ and block off-diagonal $F$. Making use of the result in \cite[][Theorem 2.3]{IPSEN} we expand the determinant as follows
\begin{equation}
\det(A+F)=\det A+\det F+S_1+S_2+\ldots +S_{2N-1},
\end{equation}
where 
\begin{equation}
S_{k}=\sum\limits_{1\leq i_1<\ldots <i_{k}\leq 2N} \frac{\det A}{a_{i_1}\ldots a_{i_k}} \det F_{i_1,\ldots,i_k}, \label{eq:Skdef}
\end{equation}
$a_{i}$ denotes the $i$-th element on the diagonal of $A$, and $F_{i_1,\ldots,i_k}$ is a $k\times k$ submatrix the element of which lie at the intersections of $i_1,\ldots,i_k$-th rows and columns.

 If $z$ is far from any of the eigenvalues of $X$, the regularization in \eqref{potKF} is not needed and one can safely set $|w|\to 0$. In the $|w|\to 0$ limit the non-vanishing contribution to the correlator comes when during the averaging procedure $z$ is close to a certain eigenvalue $\lambda_i$. 
We remark that in matrix models with unitary symmetry the probability of coalescence of two eigenvalues is zero due to the presence of the Vandermonde determinant. Since $S_1=0$, the dominant term in the expansion is 
 \begin{equation}
 S_2=\sum_{k,l=1}^{N}\frac{\det A}{(z-\lambda_k)(\zb-\bar{\lambda}_l)} \det \left(\begin{array}{cc}
0 & -\wb \<L_k|L_l\> \\
w \<R_l|R_k\> & 0  
 \end{array}\right)=\frac{\det A}{|z-\lambda_i|^2}|w|^2 O_{ii}+|w|^2\mathcal{O}(|z-\lambda_i|).
\end{equation}  
In the first equality we used the fact that $F$ is off-diagonal and the only non-zero terms correspond to $i_1 \leq N$ and $i_2\geq N$ in \eqref{eq:Skdef}. Since $\det A$ contains a factor $|z-\lambda_i|^2$, we regrouped terms into the ones that $(z-\lambda_i)$ has canceled ($k=l$), and others ($k\neq l$) which include a factor of $z-\lambda_i$ or its conjugate, denoting the sum over them by $\mathcal{O}(|z-\lambda_i|)$.
 
In the leading term we obtain
\begin{equation}
\partial_w \log\det(Q-{\cal X})=\frac{\wb}{\frac{|z-\lambda_i|^2}{O_{ii}}+|w|^2},
\end{equation}
therefore it is easy to see that
\begin{equation}
\pi O(z,\zb)=\lim_{|w|\to 0}\<\frac{1}{N^2}\partial_{w} \log\det(Q-{\cal X})\partial_{\wb}\log\det(Q-{\cal X})\>,
\label{correigenvect}
\end{equation}
where we used the representation of the two-dimensional Dirac delta \eqref{Diracdelta}. 
Equation (\ref{correigenvect}) is exact for any $N$. It is a characteristic property of probability density functions which are invariant under the action of the $U(N)$ group that in the large $N$ limit the average of two quantities $A$ and $B$ preserving the $U(N)$ symmetry factorizes.
More precisely, denote
$$f_N \equiv \frac{1}{N}\partial_{w} \log\det(Q-{\cal X}),
\qquad
g_N  \equiv 
\frac{1}{N}\partial_{\wb}\log\det(Q-{\cal X}).$$
As indicated in Section 5 of \cite{Guionnet} it is enough to prove
concentration of measure on the unitary group
(i.e., assume deterministic singular values for $X$).
Then, as in Eq. (34) in Section 3.2 of \cite{Guionnet}, by relying on
Corollary 4.4.28 of \cite{AGZ}, we have that $f_N$, $g_N$, as well as $f_Ng_N$ are almost surely
close to their respectve expected values as $N\to\infty$.

%$\<AB\>=\<A\>\<B\>+\mathcal{O}(N^{-2})$~\cite{GROSS}. 
Applying this to  \eqref{correigenvect}, we finally obtain
\begin{equation}
\pi O(z,\zb)=\lim_{|w|\to 0} \partial_{w}\Phi \partial_{\wb}\Phi=\left.-{\cal G}_{12} {\cal G}_{21}\right|_{|w|=0}. \label{CorrProduct}
\end{equation}

\section{Diagonal overlaps and the stability of the spectrum} \label{sec:DiagOverlap}
The diagonal elements of the matrix of overlaps play an important role in the stability of the spectrum of non-normal matrices as can be seen on the following example. Consider a diagonalizable matrix $X$ which is slightly perturbed by $P$: $X(\epsilon) =X+\epsilon P$. The first order perturbation yields the leading term in $\epsilon$
\begin{equation}
|\lambda_i(\epsilon)-\lambda_i(0)|=|\epsilon \<L_i|P|R_i\>|\leq |\epsilon|  \sqrt{\<L_i|L_i\>\<R_i|R_i\>} ||P||,
\end{equation}
and the bound is reached if the perturbation is of rank one $P=\left|L_i\>\<R_i\right|$. The square root of $O_{ii}$ is known in the numerical analysis community as the eigenvalue condition number $\kappa(\lambda_i)$, introduced in \cite{WILKINSON} (see also \cite{TREFETHEN} for a review). The Cauchy-Schwartz inequality asserts that $O_{ii}\geq 1$.

Generically, $O_{ii}\sim N$  in the bulk (cf. example 5 for the behavior at the edge), but for normal matrices $O_{ij}=\delta_{ij}$, showing explicitly  that the eigenvalues of normal matrices are the stablest. The definition \eqref{evs} of $O(z,\zb)$  differs from the original one in \cite{CHALKERMEHLIG} by the factor of $N^{-1}$ so that in the large $N$ limit $O(z,\zb)$ remains finite. As a consequence of this normalization, for normal matrices there holds a relation $O(z,\zb)=N^{-1}\rho(z,\zb)$.

The eigenvector correlator and the mean spectral density give a partial access to the conditional probability of the eigenvalue condition numbers, namely their ratio is the conditional mean of the eigenvalue condition number
\begin{equation}
c(z,\zb)=\mathbb{E}\left(\kappa^{2}(\lambda_i)N^{-1}|\lambda_i=z\right)=\int \frac{O_{ii}}{N}\frac{p(O_{ii},\lambda_i=z)}{p(\lambda_i=z)}dO_{ii}=\frac{1}{\rho(z,\zb)}\int \frac{O_{ii}}{N} \delta^{(2)}(z-\lambda_i)p(X)dX=\frac{O(z,\zb)}{\rho(z,\zb)}.
\end{equation}
Here we used the fact that the joint probability density for $O_{ii}$ and $\lambda_i$ can be calculated by integrating out all other variables from the joint pdf for the matrix elements. The additional $N^{-1}$ factor and the summation over the eigenvalues appears if one symmetrizes the integrand. Last, but not least, we mention that $c(z,\zb)$ is also known in physics as a Petermann factor (excess noise factor), reflecting the non-orthogonality of the cavity modes in open chaotic scattering~\cite{BEENAKKER,FYODOROVMEHLIG,BERRY}.

\section{The eigenvector correlator from the single ring theorem} \label{sec:CorrFromHL}
In the previous section we have stressed that the full solution of the spectral non-hermitian problem requires the simultaneous knowledge of its eigenvalues and eigenvectors, since they  mutually interact with each other already at the  leading terms of  the $1/N$ expansion. Considering now the case of the Haagerup-Larsen theorem, one may therefore wonder, what has happened  to the information about the correlator $O(z,\zb)$.  Certainly, $R$-diagonal operators are not normal in general, so such correlators are different from zero. Luckily, the direct construction of Belinschi, Speicher and \'{S}niady can easily give the answer, we just read off the product of the appropriate elements of the Green's function~\cite[][eq. (31)]{BSS}
\begin{equation}
{\cal G}_{12}(z,\zb,i\epsilon,-i\epsilon){\cal G}_{21}(z,\zb,i\epsilon,-i\epsilon)=\frac{\omega^2(i\epsilon)}{(\omega^2(i\epsilon)-|z|^2)^2}. \label{BSS1}
\end{equation}
The function $\omega(i\epsilon)$ is specified in~\cite{BSS}, however its exact form is not necessary for our purpose, since the upper diagonal element of the Green's function satisfies the relation~\cite[][eq. (32)]{BSS}
\begin{equation}
z G(z,\zb,i\epsilon,-i\epsilon)=\frac{|z|^2}{|z|^2-\omega^{2}(i\epsilon)} . \label{BSS2}
\end{equation}
In the limit $\epsilon \to 0$ the l.h.s. tends to $z G(z,\zb,0,0)\equiv F(|z|)$ which is the radial cumulative distribution function~\cite{SINGLERING,BSS} $F(r)=2\pi\int_{0}^{r}r' \rho(r')dr'$. It satisfies the functional equation $S_{P^2}(F(r)-1)=r^{-2}$~\cite{HAAGLAR}. For simplicity we use the notation $r=|z|$. Combining \eqref{BSS1} and \eqref{BSS2} with \eqref{CorrProduct}, we finally obtain
\be
O(r)=\frac{1}{\pi} \frac{F(r)\left[1-F(r)\right]}{r^2},
\ee
which represents the main result of this paper. 

%
%provided one  establishes the correspondence between our formulation and the notation of~\cite{BSS}. In brief, ~\cite{BSS} use so-called "hermitization" version of generalized Green's functions, corresponding to 90$^0$ rotation of basic $2$ by $2$  matrix representing the  generalized Green's function. Therefore in the formalism of ~\cite{BSS}, diagonal elements of  generalized Greens function play the role of eigenvector correlators. 
%The exact relation reads: 
%\be
%|V|^2= - \lim_{z \rightarrow i \epsilon} \frac{\omega^2(z)}{(\omega(z)^2-|\lambda|^2)^2},
%\label{vcorr_bss}
%\ee
%where on the r.h.s. we used explicitly Eq.~(31) from ~\cite{BSS}. Note, that due to the swap of diagonal with non-diagonal elements in the hermitization version of generalized 
%Green's functions, $z=i\epsilon$ variable in ~\cite{BSS}   plays the role of $w$  variable in our formulation, and $\lambda$ plays the role of our $z$.   The function $\omega(i \epsilon)$ in~\cite{BSS} practically determines the radial part of the spectrum, since Eq. (32) from~\cite{BSS} reads   ,
%\be
%F(| \lambda |)\equiv \lambda G_{\mu_x,\epsilon}(\lambda)=\frac{|\lambda|^2}{|\lambda|^2 - \omega(i\epsilon)^2}
%\label{F_bss}
%\ee
%where $F(| \lambda |)$ comes from Haagerup-Larsen construction $S_{P^2}(F(| \lambda |)-1)=\frac{1}{|\lambda|^2}$. 
%Combining (\ref{vcorr_bss}) and (\ref{F_bss}) we arrive at 
%\be
%|\lambda|^2|V|^2 =F(|\lambda|)(1-F(|\lambda|)).
%\ee
%Coming back to our notation, 
%\be
%O(r)=\frac{1}{\pi} \frac{F(r)(1-F(r))}{r^2},
%\ee
%which represents the main result of this note. 

%%%%%%%%%%%%%%%%%%%%%%%%%%%%%%%%%%%%%%%%%%

\section{Examples}
\label{examples}
In this section we provide five  examples: (1) free product of $n$ complex Ginibre ensembles,
(2) free product of $n$ truncated Haar matrices, (3) free quotient  of complex Ginibre ensembles, (4) free sum of $k$ Haar measures, (5) mean condition number for the Ginibre ensemble. 
Taking into account the simplicity of the main formula~(\ref{main}), generalizations to any domain of applicability of the single ring theorem are straightforward. 
\begin{enumerate}
\item Let us first take $Y=X_1 X_2 ...X_n$, where $X_i$ are free complex Ginibre ensembles. In this case the Haagerup-Larsen theorem yields $F_Y(r)= r^{\frac{2}{n}}$ for $r<1$ and 1 otherwise~\cite{BNS,TIKHOMIROV}, hence
\be
O_{Y,n}(r)= \frac{1-r^{2/n}}{ \pi r^{2-2/n}}\theta(1-r),
\label{proGin}
\ee 
where $\theta$ denotes the Heaviside (step) function. For completeness we note that \cite{BURDAJANIKWACLAW}
\be
\rho_{Y,n}(r)=\frac{1}{\pi n} r^{\frac{2}{n}-2}\theta(1-r).
\ee
Interestingly, even for the case $n=1$ the result for $O_{Y,1}(r)=\pi^{-1}(1-r^2)\theta(1-r)$ was obtained for the first time 33 years~\cite{JNNPZ,CHALKERMEHLIG} after the spectral density result $\rho_{Y,1}=\pi^{-1}\theta(1-r)$  derived  in the seminal paper by Ginibre~\cite{GG}.
Figure~\ref{numGin}a) confronts our  prediction with the numerical calculations. Recently, (\ref{proGin}) was confirmed by independent calculation using diagrammatic methods~\cite{BURDA}. 
\begin{figure}[ht!]
	\centering
	\includegraphics[width=0.49\textwidth]{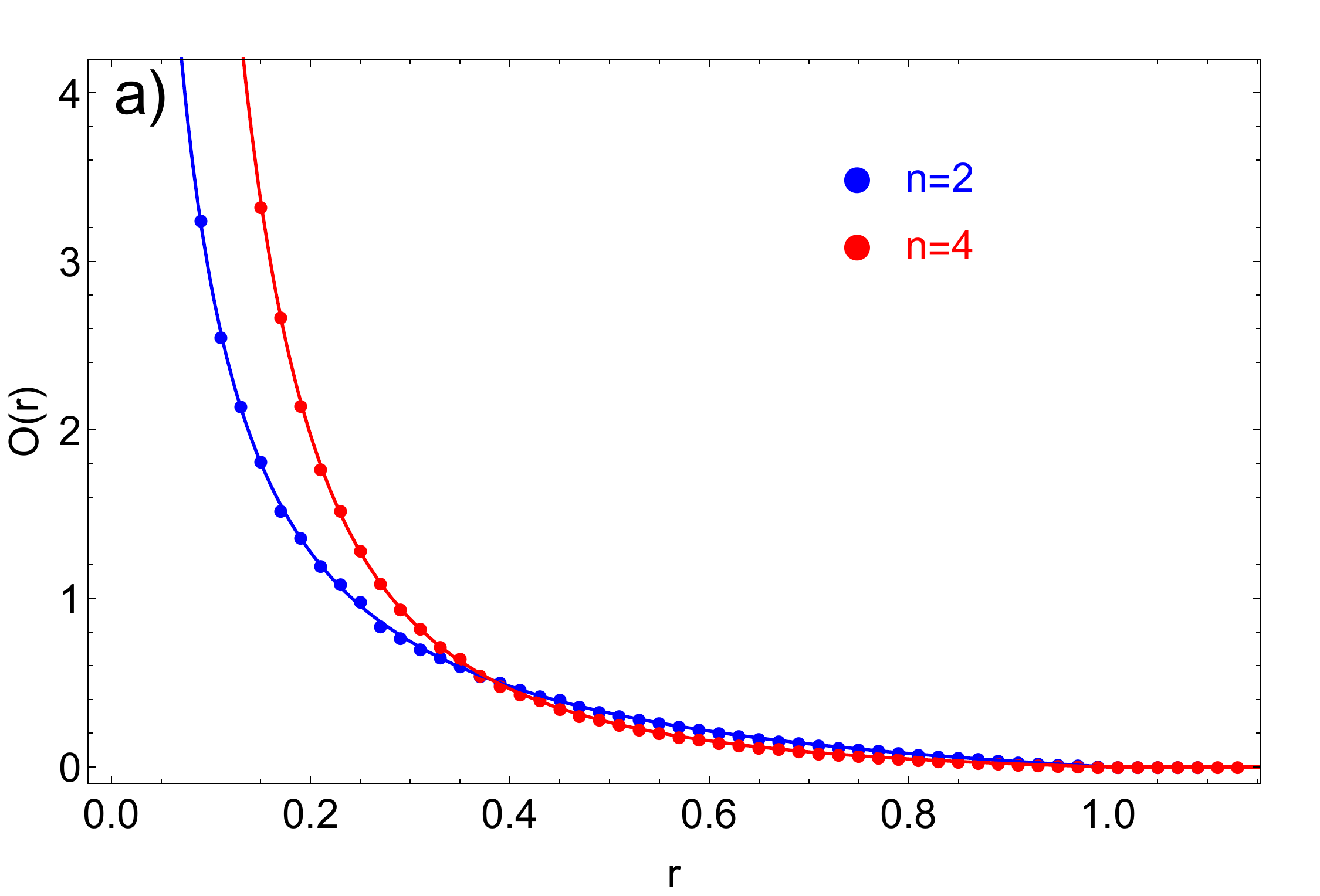}
	\includegraphics[width=0.49\textwidth]{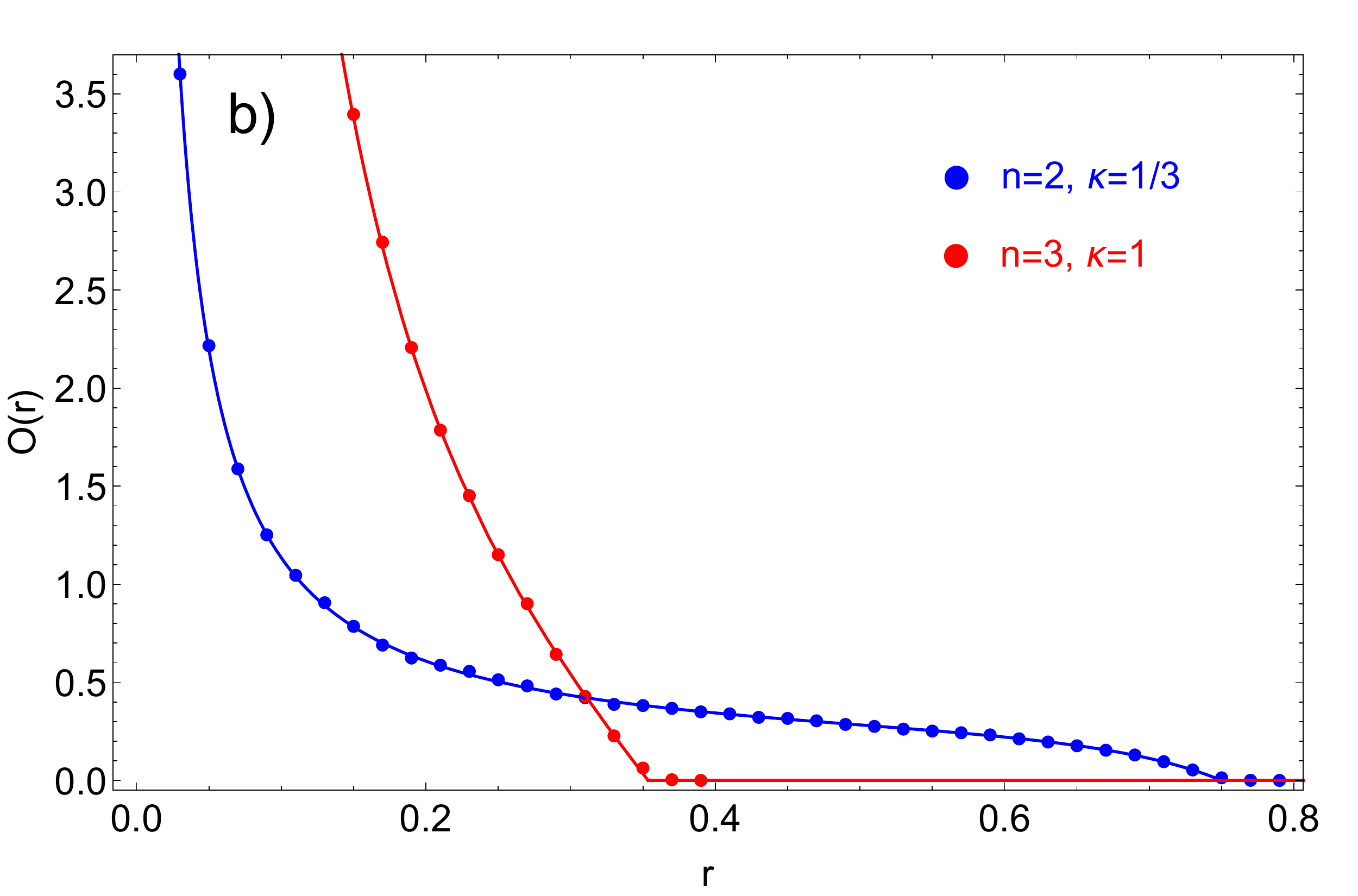}
      \caption{A numerical simulation (dots) of the eigenvector correlator for the product of a) $n=2,4$ complex Ginibre matrices of the size 1000 by 1000, averaged over the sample of 2000 matrices; b) $n=2,3$ truncated unitary  matrices of the size $L=1000$, $N=\kappa L$ done on the samples of 2000 matrices.  The solid lines represent the analytic prediction.  The fact that one observes datapoints outside the limiting spectrum is the effect of the finite size of matrices. Their agreement with the theoretical prediction ($O(r)=0$ outside) is related to the scaling of the diagonal overlap $O_{ii}$ (see also example 5).} 
      \label{numGin}
\end{figure}
\item Let us also take $Y=X_1 X_2 ...X_n$, where $X_i$ are truncated unitary matrices,  i.e.  Haar matrices of the size  $(N+L) \times (N+L)$, in which $L$ columns and rows are removed. In the limit where both $L$ and $N$ tend to infinity in such a way that $\kappa=L/N$ is fixed, 
 $F_{Y}=\kappa \frac{r^{2/n}}{1-r^{2/n}}$  for $r<(1+\kappa)^{-n/2}$ and 1 otherwise~\cite{BNS}, hence
\be
O_{Y,n,\kappa}(r)=\frac{\kappa}{\pi} \frac{1-r^{2/n}(1+\kappa)}{r^{2(n-1)/n} (1-r^{2/n})^2}\theta\left(\frac{1}{(1+\kappa)^{n/2}}-r\right).
\label{protrU}
\ee 
Figure~\ref{numGin}b) confronts our  prediction with the numerical calculations. 
%\begin{figure}[ht!]
%	\centering
%	%\includegraphics[width=1.\textwidth]{fig2.pdf}
%      \caption{A numerical simulation of the eigenvector correlator for the product of .  Red line represents the analytic prediction.}
%      \label{numUni}
%\end{figure}

\item
Let us consider $k$ Ginibre ensembles $X_{i}$ and the same number of inverse Ginibre ensembles $\tilde{X}^{-1}_{i}$. Defining $Y$ as the product $Y=X_1\ldots X_{k}\tilde{X}_{1}^{-1}\dots\tilde{X}_{k}^{-1}$, one easily obtains~\cite{HAAGSCH} $S_{P^2}(z)=(-z)^k/(1+z)^k$.  If we argue that the eigenvector correlator formula holds also for the unbounded measures, using the Haagerup-Larsen theorem we get a rather unexpected result:
\be 
\rho_{Y,k}(r)=\frac{1}{\pi k} \frac{r^{\frac{2}{k}-2}}{\left(1+r^{2/k}\right)^2}, \nonumber \\
O_{Y,k}(r)=\frac{1}{\pi r^2}\frac{r^{\frac{2}{k}}}{\left(1+r^{2/k}\right)^2}, 
\ee
i.e. the spectral density and the eigenvector correlator satisfy a very simple relation $O_{Y,k}(r)=k\rho_{Y,k}(r)$, which means that the eigenvalues are (on average) uniformly conditioned. The formula for the spectral density agrees with the recent results~\cite{ZENGSPHERICAL}. In Fig. \ref{fig:GGinvGG}a) we confront our result for the eigenvector correlator with numerical simulations.

\begin{figure}
\includegraphics[width=0.49\textwidth]{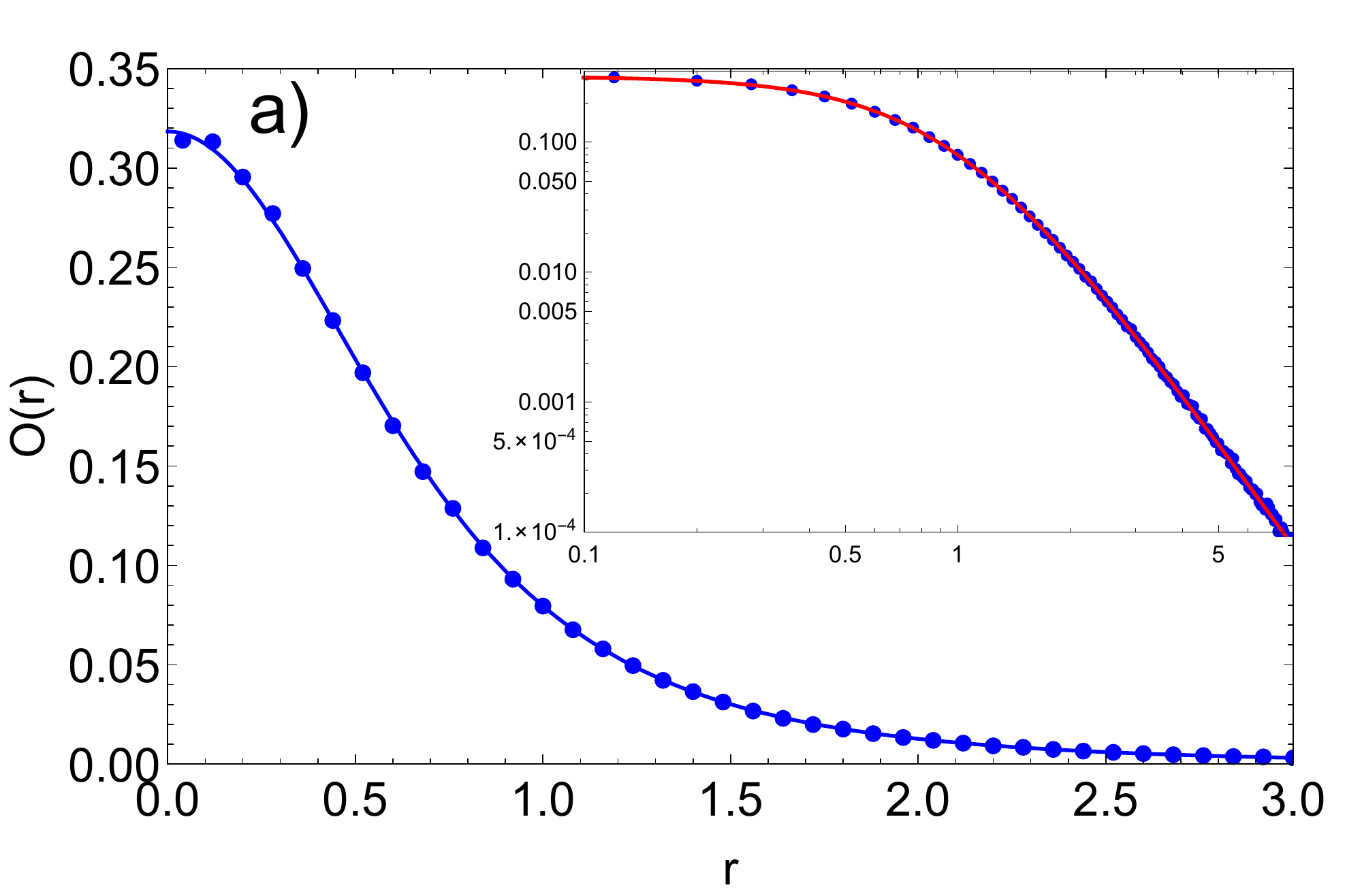} 
\includegraphics[width=0.49\textwidth]{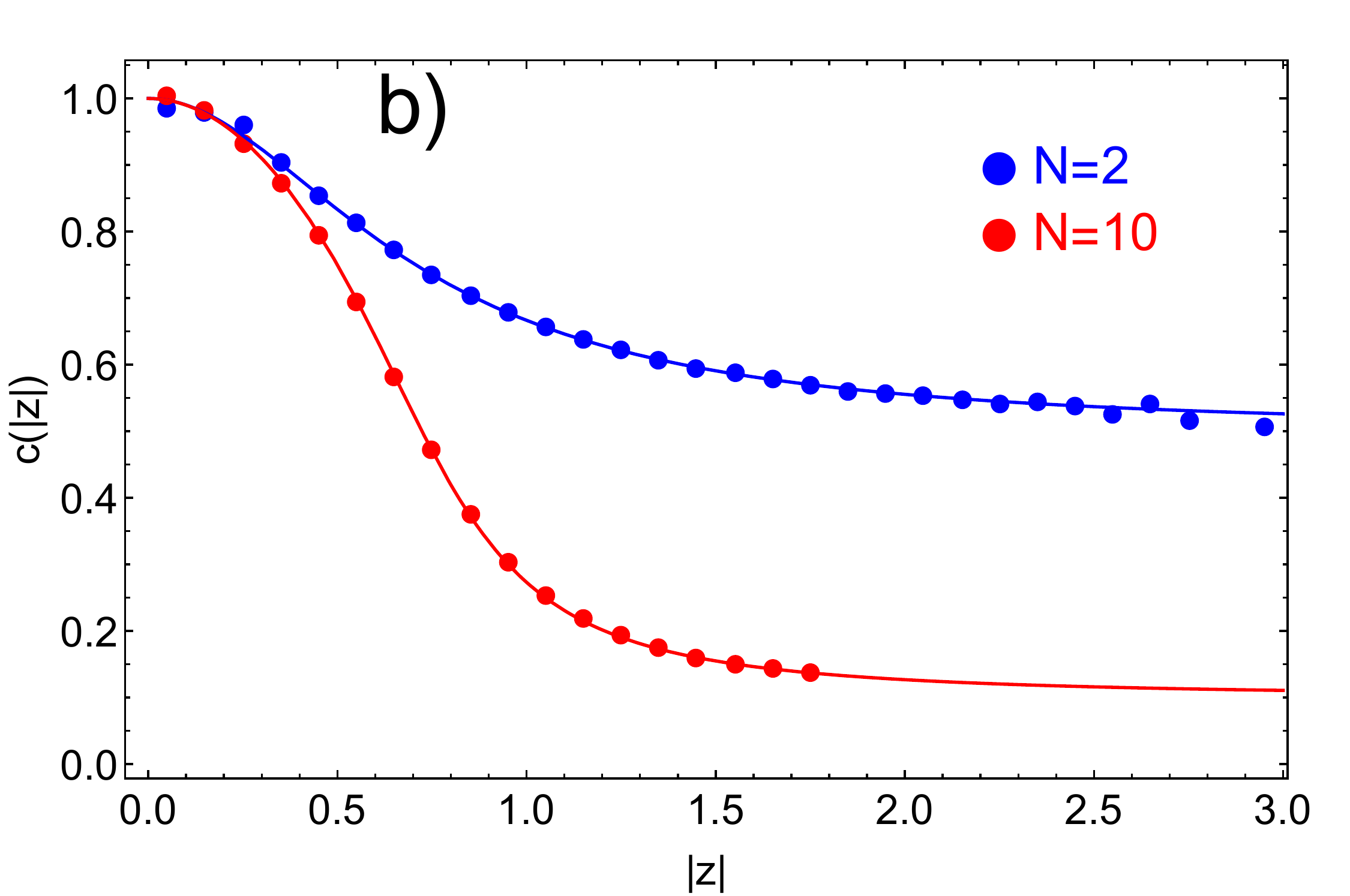}
\caption{a) The eigenvector correlator calculated by a numerical diagonalization of 4000 matrices that are ratio of two Ginibres (also known as the spherical ensemble) of size $N=1000$ presented on linear and double logarithmic (inset) scales. b) Mean eigenvalue condition number of the Ginibre ensemble. The complex plane was divided into the hollowed cylinders of radii $r$ and $r+\Delta r$, eigenvalues and their condition numbers were assigned to cylinders, according to the modulus of the eigenvalue. The dots denote the average eigenvalue condition number within each cylinder, the lines present formula \eqref{eq:GinibreCondNum}. Numerical distribution was obtained by the diagonalization of $10^6$ matrices of size $N=2$ and $4\cdot 10^5$ matrices of size $N=10$.
\label{fig:GGinvGG}}
\end{figure}

%
%Let Y be a free quotient of complex Ginibre ensembles $X_1,X_2$: $Y=X_1 X_2^{-1}$. It is well known that such quotient gives the squared Cauchy type spectral distribution~\cite{HAAGSCH}, or equivalently,
%$F_Y(r)=\frac{r^2}{1+r^2}$. If we argue, that the eigenvector correlator formula holds also for the unbounded measures, we obtain a rather unexpected result
%\be
%\rho_Y(r)&=&\frac{1}{2 \pi r} \frac{dF_Y(r)}{dr}=\frac{1}{\pi}\frac{1}{(1+r^2)^2} \nonumber \\
%O_Y(r) &=& \frac{1}{\pi} \frac{F_Y(r)(1-F_Y(r))}{r^2}=\frac{1}{\pi}\frac{1}{(1+r^2)^2} ,
%\ee
%i.e. spectral density and eigenvector correlator are given by the same formula. Further generalization to the product of $k$ mutually free complex Ginibre matrices $X_i$ with $k$ inverse Ginibre matrices $\tilde{X}_{i}^{-1}$: $Y=X_1\ldots X_{k}\tilde{X}_{1}^{-1}\dots\tilde{X}_{k}^{-1}$ is straightforward  and then
%\begin{equation}
%O_{Y,k}(r)=\frac{r^{2/k}}{\pi (1+r^{2/k})^2}.
%\end{equation}
% The relation between the spectral density and the correlator is only slightly modified: $O_{Y,k}(r)=k \rho_{Y,k}(r)$.
\item 
Let us consider free convolution of $k$  Haar-distributed  matrices $U_k$, i.e. $Y=U_1+U_2+\dots+U_k$. Then, $F_Y=\frac{r^2(k-1)}{k^2-r^2}$~\cite{HAAGLAR,JAROSZ} for $r <\sqrt{k}$ and 1 otherwise, hence
\be
\rho_{Y,k}(r)&=&\frac{1}{\pi} \frac{k^2(k-1)}{(k^2-r^2)^2}\theta(\sqrt{k}-r), \nonumber \\
O_{Y,k}(r) &=&\left(1-\frac{r^2}{k}\right) \rho_{Y,k}(r). \label{eq:UnitSumCorr}
\ee

%Comparison of this result with numerical simulations is presented in Fig. \ref{fig2}.
%
%\begin{figure}[ht]
%\begin{center}
%\includegraphics[width=0.49\textwidth]{corrsumU_2_4.pdf}
%\end{center}
%\caption{A numerical simulation (dots) of the eigenvector correlator of the sum of $n=2,4$ Haar unitary matrices of size 1000 by 1000 with 2000 independent realizations. Solid lines represent analytic results. 
%\label{fig2} }
%\end{figure}

\item  Our formula \eqref{main} is valid only in the limit $N\to\infty$. In order to access condition numbers in the Ginibre ensemble in the finite $N$, we superimpose the results from~\cite{CHALKERMEHLIG} and~\cite{STARR}, derived with the use of different techniques. We obtain the formula for the averaged  squared eigenvalue condition number
\begin{equation}
c(z,\zb)=1-|z|^2+\frac{1}{N}\frac{e^{-N|z|^2}(N|z|^2)^N}{\Gamma(N,N|z|^2)},
\label{eq:GinibreCondNum}
\end{equation}
valid for any size of a matrix. Here $\Gamma(N,x)$ stands for the incomplete gamma function. The accordance with the numerical results is presented in Fig. \ref{fig:GGinvGG}b).

Formulas \eqref{proGin} \eqref{protrU} and \eqref{eq:UnitSumCorr} predict that the eigenvector correlator vanishes at the edges of the spectrum. It  neither means that the eigenvectors become orthogonal to each other nor that condition number is smaller than 1, rather the diagonal overlap grows slower than linearly  with the size of a matrix. Indeed, an asymptotic analysis of the formula \eqref{eq:GinibreCondNum} yields the conditional expectation 
\begin{equation}
\mathbb{E}(O_{ii}|1=|\lambda_i|)= \sqrt{\frac{2}{\pi}}N^{1/2} +\frac{2}{3\pi} +\mathcal{O}(N^{-1/2})
\end{equation} 
\end{enumerate}

%%%%%%%%%%%%%%%%%%%%%%%%%%%%%%%%%%%%%%%%%%%%%%%%%%%%%%%%%%%%%%%%%%%%%%
%%%%%%%%%%%%%%%%%%%%%%%%%%%%%%%%%%%%%%%%%%%%%%%%%%%%%%%%%%%%%%%%%%%%%%%%%%%%%%%%%%%%%%%%%%%%%%%%%%%%%%%%%%%%%%%%%%%%%%%%%
\section{Eigenvector-eigenvalue mapping} \label{sec:eigenvec-eigenval}
Combining (\ref{main}) with the definition   of the radial spectral density $\rho(r)=\frac{1}{2 \pi r} \frac{dF(r)}{dr}$, we can  obtain a general  relation between $\rho(r)$ and $O(r)$. Solving a quadratic equation  we arrive at 
\be
F(r)=\frac{1}{2} ( 1 \mp \sqrt{1-4\pi r^2 O(r)}).
\label{quadratic} 
\ee
We remark that both signs are relevant. At the inner rim of the ring, $F(r)=0$, which corresponds to the negative sign in (\ref{quadratic}).  At the outer rim of the ring, $F(r)=1$, corresponding to the positive sign in (\ref{quadratic}).  Differentiation of (\ref{quadratic}) with respect to $r$ yields
\be
\rho(r)=\pm \frac{1}{2\sqrt{1-4 \pi r^2 O(r)}}\left(     2O(r) +r \frac{dO(r)}{dr}\right).
\label{relation}
\ee 
It is instructive to consider the Ginibre case. 
The spectral density calculated from (\ref{relation}), with the use of $O(r)=\frac{1}{\pi} (1-r^2)\theta(1-r)$,  reads 
\be
\rho(r)=\pm \frac{1}{\pi} {\rm sgn} (1-2r^2)\theta(1-r)=\frac{1}{\pi}\theta(1-r),
\ee
where the switch from the branches of  the square root in (\ref{quadratic}) takes place at $r=1/\sqrt{2}$. 
\section{Conclusions}
\label{conclusions}
We have augmented the single ring theorem with the additional prediction for certain eigenvector correlations. We pointed out a link between the main object of this paper and the sensitivity of eigenvalues to perturbations. The considered correlation function, which is the spectral density weighted by the squared eigenvalue condition number, gives partial access to the distribution of eigenvalue condition numbers. We have shown that the ratio of the eigenvector correlation function and the spectral density gives the conditional expectation of the squared eigenvalue condition number. This ratio varies on the complex plane, indicating that the eigenvalues are not uniformly conditioned.

In a series of recent papers~\cite{US,OLEG} it was argued  that the consistent description of non-hermitian ensembles requires the knowledge of the detailed dynamics of the co-evolving eigenvalues and eigenvectors. Unexpectedly,  in the Gaussian case (Ginibre ensemble),  the dynamics of eigenvectors seemed to play even a superior role (at least in the $N \to \infty$ limit) and leads directly to the inference of the spectral properties solely from the knowledge of the eigenvector correlator. 
The case considered here, the Haagerup-Larsen theorem, seems to agree with this scenario. 
We conjecture therefore that in generic non-hermitian ensembles the correlations between left and right eigenvectors play an equally important role as  the spectral information. Historically, in the literature on non-hermitian random matrix models, our parameter $w$ always played the role of a regulator of the Fuglede-Kadison determinant, and was usually put to zero in an incautious way, and thus loosing a track of eigenvectors. That practice was a consequence of the duplication  of the  paradigm of hermitian random matrix models, which concentrates on the spectrum, since the $U(N)$ invariance of the probability density function leads to the decoupling of the eigenvectors. We suggest that this paradigm has to be challenged in the case of non-normal random matrix models, where the unitary transformations are not sufficient to diagonalize a matrix. In particular, our  formula~\eqref{quatgf} explicitly points at the symmetric nature of $z$ and $w$ as complex variables, controlling the spectra and eigenvectors, respectively. In particular, $\partial_w {\cal G}_{11}=\partial_z {\cal G}_{12}$. 

The presented new relation for eigenvectors in the single ring theorem is just the consequence  of the above-mentioned  symmetry applied to $R$-diagonal operators. 

On more general grounds, it is tempting to speculate that the interplay between eigenvector correlators and spectral measures may play a role in 
generalizations of the Brown measure, that will be free of pathological discontinuities, as observed in~\cite{SNIADY}.

%%%%%%%%%%%%%%%%%%%%%%%%%%%%%%%%%%%%%%%%%%%%%%%%%%%%%%%%%%%%%%%%%%%%%%
%%%%%%%%%%%%%%%%%%%%%%%%%%%%%%%%%%%%%%%%%%%%%%%%%%%%%%%%%%%%%%%%%%%%%%
\section*{Acknowledgments}
MAN and WT were supported by the Grant DEC-2011/02/A/ST1/00119 of the National Centre of Science. WT
appreciates also the support from the Ministry of Science and Higher Education through
the Diamond Grant 0225/DIA/2015/44. 
The work of RS was supported by the ERC Advanced Grant NCDFP 339760.
We are grateful to Jacek Grela, Ewa Gudowska-Nowak, Romuald A. Janik   and Piotr Warcho\l{} for discussions.

%%%%%%%%%%%%%%%%%%%%%%%%%%%%%%%%%%%%%%%%%%%%%%%%%%%%%%%%%%%%%%%%%%%%%%
%%%%%%%%%%%%%%%%%%%%%%%%%%%%%%%%%%%%%%%%%%%%%%%%%%%%%%%%%%%%%%%%%%%%%%

%%%%%%%%%%%%%%%%%%%%%%%%%%%%%%%%%%%%%%%%%%%%%%%%%%%%%%%%%%%%%%%%%%%%%%
%%%%%%%%%%%%%%%%%%%%%%%%%%%%%%%%%%%%%%%%%%%%%%%%%%%%%%%%%%%%%%%%%%%%%%

%%%%%%%%%%%%%%%%%%%%%%%%%%%%%%%%%%%%%%%%%%%%%%%%%%%%%%%%%%%%%%%%%%%%%%

%%%%%%%%%%%%%%%%%%%%%%%%%%%%%%%%%%%%%%%%%%%%%%%%%%%%%%%%%%%%%%%%%%%%%%

%%%%%%%%%%%%%%%%%%%%%%%%%%%%%%%%%%%%%%%%%%%%%%%%%%%%%%%%%%%%%%%%%%%%%%
%%%%%%%%%%%%%%%%%%%%%%%%%%%%%%%%%%%%%%%%%%%%%%%%%%%%%%%%%%%%%%%%%%%%%%

%%%%%%%%%%%%%%%%%%%%%%%%%%%%%%%%%%%%%%%%%%%%%%%%%%%%%%%%%%%%%%%%%%%%%%
%%%%%%%%%%%%%%%%%%%%%%%%%%%%%%%%%%%%%%%%%%%%%%%%%%%%%%%%%%%%%%%%%%%%%%

\end{document}